\documentclass[superscriptaddress,prd,aps,showpacs,nofootinbib,showkeywords,eqsecnum,preprint]{revtex4-2}

\usepackage{graphicx,color,amsmath,amsxtra}
\usepackage{epsf}
\usepackage{amssymb}
\usepackage{enumerate}
\usepackage{hhline}
\usepackage{array}
\usepackage{tabularx}
\usepackage[unicode]{hyperref}

\begin{document}
\pagestyle{myheadings}
\title{Stability analysis of de Sitter solution in the Starobinsky-Grisaru-Zanon gravity using the dynamical system method}
\author{Tuan Q. Do}
\email{tuan.doquoc@phenikaa-uni.edu.vn}
\affiliation{Phenikaa Institute for Advanced Study, Phenikaa University, Hanoi 12116, Vietnam}

\begin{abstract}
We study whether the so-called Starobinsky-Grisaru-Zanon gravity, which is a novel fourth-order gravity model involving the so-called Grisaru-Zanon term, admits a stable de Sitter solution.  First, we derive the corresponding field equations of the Starobinsky-Grisaru-Zanon gravity for the spatially flat Friedmann-Lemaitre-Robertson-Walker metric by using an effective method based on the Euler-Lagrange equations. Then, we figure out an exact de Sitter solution of these field equations for the first time.  Finally,  we point out, through the dynamical system method, that the obtained de Sitter solution is always unstable. Interestingly, although  the Starobinsky $R^2$ term does not contribute to the value of the obtained de Sitter solution, it does affect on the instability of this solution. 
\end{abstract}
\maketitle
\newpage
\section{Introduction} \label{intro}
After four decades,  a fourth-order gravity model proposed by Starobinsky \cite{Starobinsky:1980te}, which is one of the very first inflationary models, along with models by Guth \cite{Guth:1980zm}, Linde \cite{Linde:1981mu,Linde:1983gd}, and few others, has still been one of the most viable models in the light of the latest data of the Planck satellite constructed to probe the cosmic microwave background radiations (CMB) \cite{Akrami:2018odb}. The cosmological viability of the Starobinsky inflationary model (or the Starobinsky model for short) is basically based on the including of quadratic Ricci scalar term, i.e., $R^2$,  into the pure Einstein-Hilbert action \cite{Starobinsky:1980te}. Theoretically, this term can be regarded, according to Starobinsky, as a quantum correction to the Einstein's gravity.

In addition to the succeed in inflationary predictions, the Starobinsky model has several other interesting and unique aspects, which can be listed as follows.  First, it appears as  the simplest model of the class of fourth-order gravity,  whose cosmological implication is very rich, not only for an early universe \cite{Whitt:1984pd,Maeda:1987xf,Barrow:1988xh}, but also for a late-time universe \cite{Carroll:2004de,Nojiri:2010wj,Nojiri:2017ncd}.   Besides, it is expected that a renormalizable model of gravity may be constructed if higher-order curvature correction terms, e.g., $R^2$ and $R_{\mu\nu}R^{\mu\nu}$, were introduced into the pure Einstein-Hilbert action \cite{Stelle:1976gc}. Second, the Starobinsky model is a rare fourth-order gravity model, which is free of the so-called Ostrogradsky ghost expected to emerge from higher-order derivatives  existing in field equations \cite{Woodard:2015zca}. To have a wider view of fourth-order gravity, see interesting reviews in Refs. \cite{Schmidt:2006jt,Alvarez-Gaume:2015rwa,Salvio:2018crh}. On the legacy of the Starobinsky model, see Ref. \cite{Ketov:2025nkr}.

Despite these important points, the cosmological validity of the Starobinsky model might still be challenged by more precise CMB observations. For example, a very recent data of the Atacama Cosmology Telescope (ACT) has pointed out that the prediction of the Starobinsky model seems to be slightly disfavored \cite{AtacamaCosmologyTelescope:2025nti}. Hence, non-trivial extensions of the Starobinsky model could be necessary in order to have more viable inflationary model(s). Interestingly, this task has been done extensively, even before the release of ACT data, e.g., see Refs. \cite{Myrzakulov:2014hca,Cano:2020oaa,Rodrigues-da-Silva:2021jab,Ivanov:2021chn,Gialamas:2023lxj,Modak:2022gol,Ketov:2022lhx,Ketov:2022zhp,Do:2023yvg,Pham:2024fub,Toyama:2024ugg,Asorey:2024oxw} for recent relevant examples. In addition, see Refs. \cite{Addazi:2025qra,Bianchi:2025tyl,Ketov:2025cqg} for updates related to the data of ACT within the context of the Starobinsky model. 

Among these extensions, we would like to extend our current interest in fourth-order gravities to the so-called Starobinsky-Grisaru-Zanon (SGZ) gravity model, which has been proposed recently in a paper \cite{Toyama:2024ugg}. As a result, this paper is a follow-up study of the so-called Einstein-Grisaru-Zanon (EGZ) gravity model considered previously in Ref. \cite{CamposDelgado:2024jst}. In these two models, there exists the so-called Grisaru-Zanon (GZ) term, which is the leading superstring correction term  firstly pointed out  by Grisaru and Zanon in Ref. \cite{Grisaru:1986vi}. 

Similar to our previous study of the EGZ gravity \cite{preprint}, the mathematical tools, which we will use to investigate whether the SGZ gravity admits a stable de Sitter solution, are the Euler-Lagrange (EL) equations as well as the dynamical system \cite{Bahamonde:2017ize}. These methods have been successfully used in our other papers \cite{Do:2023yvg,Pham:2024fub,Do:2020vdc}. It seems to us that these methods have appeared as effective methods so far, especially when we deal with very complicated fourth-order gravities like the EGZ and SGZ models. 

Before going to present explicitly our results in the following sections, we would like to note that the existence and stability of de Sitter solution can play an important criterion for determining whether the SGZ gravity could potentially be a realistic inflationary model, according to the discussions in Refs. \cite{Elizalde:2014xva,Pozdeeva:2019agu,Vernov:2021hxo} (see also Refs. \cite{Toporensky:2006kc,Kamenshchik:2024kay} for related works). In particular, if the SGZ gravity admits an unstable de Sitter solution or no de Sitter solution, it would not face to the so-called eternal inflation issue, which could lead to a multiverse scenario \cite{Guth:2007ng}. Hence, it could be relevant to describe an inflationary phase of the early universe. In such a case, the so-called graceful exit mechanism \cite{Brustein:1994kw} might be unnecessary. On the other hand, if the SGZ gravity admits a stable de Sitter solution, it would only be compatible to an accelerated expansion of the late-time universe. It turns out that this important issue has not been addressed in the original paper of the SGZ gravity \cite{Toyama:2024ugg}. 

For convenience, we would like to highlight the main results contained in the present paper. First, an exact analytical de Sitter solution for the SGZ gravity is found by solving its EL equations. Second, the obtained de Sitter solution is proved to be generically unstable against field perturbations. This point serves as an important criterion for establishing the SGZ gravity as a non-trivial candidate for a realistic inflationary model. In other words, the SGZ gravity can now be safely viewed as a promising  modification of the Starobinsky model, that might help us to resolve its tension with the latest ACT data \cite{AtacamaCosmologyTelescope:2025nti,Addazi:2025qra,Bianchi:2025tyl,Ketov:2025cqg}.

The present paper will be organized as follows: (i) Its brief introduction has been written in Sec. \ref{intro}. (ii) Basic setup of the SGZ gravity will be presented in Sec. \ref{sec2}. (iii) Exact de Sitter solution of the SGZ gravity will be solved in Sec. \ref{sec3}. (iv) Stability analysis of the  obtained de Sitter solution will be investigated explicitly by using the dynamical system method in Sec. \ref{sec4}. (v) Finally, main results of the present paper will be concluded in Sec. \ref{final}.  
\section{Basic setup} \label{sec2}
\subsection{Action of Starobinsky-Grisaru-Zanon gravity}
In order to have a comparison between the EGZ and SGZ gravities, we would like to begin this section by recalling an action of the EGZ gravity considered in Ref. \cite{CamposDelgado:2024jst}, which is given by
\begin{equation}
S_{\rm EGZ}= \frac{M_p^2}{2}\int d^4 x \sqrt{-g}  \left(R+ \frac{\gamma}{M_p^6} J \right),
\end{equation}
where $M_p\equiv 1/\sqrt{8\pi G}$ is the reduced Planck mass, $\gamma$ is a dimensionless coupling constant, and $J$ is the GZ term, which is nothing but the leading superstring correction firstly worked out by Grisaru and Zanon in Ref. \cite{Grisaru:1986vi}, 
\begin{equation}
J = \left( R^{\mu\rho\sigma\nu} R_{\lambda\rho\sigma\tau} +\frac{1}{2} R^{\mu\nu\rho\sigma}R_{\lambda \tau \rho\sigma}\right)R_\mu{}^{\alpha \beta \lambda}R^\tau{}_{\alpha\beta\nu}.
\end{equation}
 It is worth noting that $\gamma$ should be positive definite in order to have the corresponding de Sitter solution for the EGZ gravity as pointed out in Refs. \cite{CamposDelgado:2024jst,preprint}. Furthermore, the positivity of $\gamma$ leads to consistency with the well-known result that the Hawking temperature of black holes decreases in superstring gravity as pointed out in Ref. \cite{CamposDelgado:2024jst}. Generally, the value of $\gamma$ is due to a compactification from ten to four dimensions and the unknown vacuum expectation value of string dilaton. However, an upper bound on $\gamma$ such as $\gamma < 1.62 \times 10^{-5}$ has been worked out in Ref. \cite{CamposDelgado:2024jst} from the Hawking temperature. 

On the other hand, an action of the SGZ gravity considered in the follow-up paper \cite{Toyama:2024ugg} reads 
\begin{equation}
S_{\rm SGZ}= \frac{M_p^2}{2}\int d^4 x \sqrt{-g}  \left(R+\frac{1}{6M^2}R^2 +\frac{\gamma} {M^6} J \right),
\end{equation}
where $R^2$ is nothing but the well-known Starobinsky term coming from the seminal paper on cosmic inflation \cite{Starobinsky:1980te}. In addition, $M$ is an additional parameter, which is normally identified with the inflaton mass (a.k.a. the scalaron mass). It is noted that the parameter $\gamma$ has been rescaled,  in the presence of $R^2$, such as $\gamma \to  \gamma \left(M_p/M\right)^6$.  Remarkably, another upper bound of $\gamma$ has been worked out in Ref. \cite{Toyama:2024ugg},  from demanding the absence of negative energy fluxes (or unitarity and causality constraints) \cite{Cano:2020oaa}, to be $\gamma <1.12 \times 10^{-6}$. For now, one can conclude that $\gamma$ must be much smaller than one in both EGZ and SBZ gravities. 

For convenience, we will rewrite the action of the SGZ gravity such as
\begin{equation}
S_{\rm SGZ}= \frac{M_p^2}{2}\int d^4 x \sqrt{-g}  \left(R+\hat\gamma R^2 +\bar\gamma J \right),
\end{equation}
where two additional parameters have been introduced,
\begin{equation} \label{def-gamma}
\hat\gamma = \frac{1}{6M^2} >0, \quad \bar\gamma = \frac{\gamma} {M^6}.
\end{equation}
In the rest of this paper, we will explore whether the SGZ gravity admits (un)stable de Sitter solutions. 
\subsection{Field equations}
In order to seek a de Sitter solution to the SGZ gravity, we need to consider the following spatially flat Friedmann-Lemaitre-Robertson-Walker (FLRW) metric, similar to our previous investigation on the EGZ gravity \cite{preprint},
\begin{equation} \label{metric}
ds^2 =-N^2(t)dt^2 +e^{2\alpha(t)}dx^2 + e^{2\alpha(t)} \left(dy^2 +dz^2 \right).
\end{equation}
Here, $N(t)$ is the lapse function, whose existence is vital for deriving the Friedmann equation from its corresponding Euler-Lagrange equation \cite{Myrzakulov:2014hca,Do:2023yvg,Pham:2024fub,Asorey:2024oxw,preprint,Do:2020vdc,Toporensky:2006kc,Kao:1991zz}. In addition, $\alpha(t)$ can be regarded as the scale factor, whose value will tell us an evolution of the universe. It is known that the lapse function $N(t)$ should be set to be one in a final version of field equations \cite{Myrzakulov:2014hca,Do:2023yvg,Pham:2024fub,Asorey:2024oxw,preprint,Do:2020vdc,Toporensky:2006kc,Kao:1991zz}. In other words, the finalized field equations of the SGZ gravity should involve only $\alpha(t)$ and its time derivatives.  It is worth noting that clear explanations for setting $N(t)=1$ have been given in Ref. \cite{Asorey:2024oxw}. Accordingly, $N(t)=1$ seems to be the simplest choice compatible with the Einstein's general relativity and its diffeomorphism invariance.

To figure out the corresponding field equations of the SGZ gravity for the FLRW metric, we continue to use an effective calculation approach based on the EL equations, following our previous study on the EGZ gravity \cite{preprint}. See also Refs.  \cite{Myrzakulov:2014hca,Do:2023yvg,Pham:2024fub,Do:2020vdc}, especially a very recent paper \cite{Asorey:2024oxw} by other people, for the use of the EL equations in deriving field equations of other fourth-order gravities. 

Before considering the EL equations, the first thing we need to do is to determine the corresponding Lagrangian of the EGZ gravity, which is defined as
 \begin{equation}
 {\cal L} =\sqrt{-g} \left(R+\hat\gamma R^2 +\bar\gamma J \right),
 \end{equation}
 for the FLRW metric. As a result, it appears that \cite{preprint}
\begin{align} \label{L1}
 R  \equiv g^{\mu\nu} R_{\mu\nu}= -6N^{-2} \left(N^{-1} \dot N \dot\alpha -\ddot\alpha-2\dot\alpha^2 \right), \quad  \sqrt{-g}  \equiv  \sqrt{-\left(\det{g_{\mu\nu}} \right)} = N e^{3\alpha}, 
 \end{align}
 where $\dot\alpha \equiv d\alpha/dt$ as well as $\ddot\alpha \equiv d^2\alpha/dt^2$.  Additionally, the corresponding value of the GZ term turns out to be \cite{preprint}
 \begin{equation} \label{L2}
J = \frac{6}{N^{12}}\left(j_4 N^4 + j_3N^3 +j_2N^2 +j_1 N+j_0 \right),
\end{equation}
with
\begin{align}
j_4&= \ddot\alpha^4+8\dot\alpha^2 \ddot\alpha^3 +22 \dot\alpha^4 \ddot\alpha^2 + 24\dot\alpha^6 \ddot\alpha+12\dot\alpha^8,\\
j_3&= -4\dot N \dot\alpha \left(\ddot\alpha^3 +6\dot\alpha^2 \ddot\alpha^2 +11 \dot\alpha^4 \ddot\alpha +6\dot\alpha^6 \right),\\
j_2&= 2\dot N^2 \dot\alpha^2 \left( 3\ddot\alpha^2 +12 \dot\alpha^2 \ddot\alpha +11\dot\alpha^4 \right),\\
j_1&= -4\dot N^3 \dot\alpha^3 \left(\ddot\alpha +2\dot\alpha^2 \right),\\
j_0 &= \dot N^4 \dot\alpha^4.
\end{align}
Apparently, $J \to 72Z $ in the limit $N(t)\to 1$, with $Z$ being defined in Ref. \cite{Toyama:2024ugg}.

According to these definitions,  it now becomes clear  that ${\cal L}$,  the Lagrangian of the SGZ gravity, is a functional of not only the first-order time derivative of $N(t)$ but also the second-order time derivative of $\alpha(t)$. Therefore, we will have two EL equations for these two variables. The first one is the EL equation for the lapse function, $N(t)$, which reads
\begin{align}
\frac{\partial {\cal L}}{\partial N} -\frac{d}{dt} \left(\frac{\partial {\cal L}}{\partial \dot N}\right)=0,
\end{align}
whose explicit expression turns out, after setting $N(t)=1$, to be
\begin{align} \label{equation-1-FLRW}
&\dot \alpha ^2 + 6\hat\gamma \left(2\dot\alpha \alpha^{(3)} -\ddot\alpha^2 +6\dot\alpha^2 \ddot\alpha \right) \nonumber\\
&+ \bar\gamma \left[ 4 \dot\alpha \left(3\ddot\alpha^2 + 12 \dot\alpha^2 \ddot\alpha +11 \dot\alpha^4 \right)  \alpha^{(3)}  - 3 \ddot\alpha^4 + 28 \dot\alpha^2 \ddot\alpha^3 + 138 \dot\alpha^4 \ddot\alpha^2 + 132 \dot\alpha^6 \ddot\alpha -12  \dot\alpha^8 \right]  =0.
\end{align}
It is worth noting that another version of this equation with $H\equiv \dot\alpha$, $\dot H \equiv \ddot\alpha$, and $\ddot H \equiv \alpha^{(3)}$ can be found in Ref. \cite{Toyama:2024ugg}. On the other hand, the last one is nothing but  the EL equation for the scale factor, $\alpha(t)$, whose mathematical formula takes
\begin{align}
\frac{\partial {\cal L}}{\partial \alpha} -\frac{d}{dt} \left(\frac{\partial {\cal L}}{\partial \dot \alpha}\right) + \frac{d^2}{dt^2} \left(\frac{\partial {\cal L}}{\partial \ddot\alpha}\right)&=0,
\end{align}
which will arrive, also after setting $N(t)=1$, at
\begin{align}\label{equation-2-FLRW}
&2\ddot\alpha+ 3\dot\alpha^2 + 6\hat\gamma \left(2\alpha^{(4)} +12\dot\alpha \alpha^{(3)} +9\ddot\alpha^2 +18\dot\alpha^2 \ddot\alpha \right) \nonumber\\
&+ \bar\gamma \left[ 4 \left(3\ddot\alpha^2 + 12 \dot\alpha^2 \ddot\alpha +11 \dot\alpha^4 \right) \alpha^{(4)} + 24 \left( \ddot\alpha +2 \dot\alpha^2 \right) \left(\alpha^{(3)} \right)^2  \right.\nonumber\\
&\left. +  \dot\alpha \left( 264 \ddot\alpha^2 +640 \dot\alpha^2 \ddot\alpha +264 \dot\alpha^4 \right)  \alpha^{(3)} +47 \ddot \alpha^4 +636 \dot\alpha^2 \ddot\alpha^3 +1206 \dot\alpha^4 \ddot\alpha^2 + 300 \dot\alpha^6 \ddot\alpha  -36 \dot\alpha^8  \right] =0.
\end{align}
It is apparent that two Eqs. \eqref{equation-1-FLRW} and \eqref{equation-2-FLRW} are indeed our desired field equations for the SGZ gravity. In these equations, the notation $\alpha^{(n)} \equiv d^n \alpha/dt^n$ stands for the $n$th-order time derivative of $\alpha(t)$. Mathematically, the field equations of SGZ gravity are indeed higher-order ordinary differential equations (ODEs). In particular,  Eq. \eqref{equation-2-FLRW} is a fourth-order ODE, while Eq. \eqref{equation-1-FLRW} is a third-order ODE. Hence, the SGZ gravity can be classified as a fourth-order gravity model, similar to the EGZ gravity \cite{preprint}. Furthermore, it is straightforward to check that two field equations, i.e., Eqs. \eqref{equation-1-FLRW} and \eqref{equation-2-FLRW}, will reduce to that defined in our previous paper \cite{preprint} once the  limit $\hat\gamma \to 0$ is taken. To end this section, we would like to emphasize that Eq. \eqref{equation-2-FLRW} can be resulted from a suitable combination of Eq. \eqref{equation-1-FLRW} and its time derivative. In other words, Eq. \eqref{equation-2-FLRW}  can be regarded as a differential consequence of Eq. \eqref{equation-1-FLRW}, in harmony with the well-known Bianchi identity $\nabla^\mu G_{\mu\nu}=0$ with $G_{\mu\nu} \equiv R_{\mu\nu}-\frac{1}{2}R g_{\mu\nu}$ being the well-known Einstein tensor and $\nabla_\mu$ is the covariant derivative. This point indicates that Eqs. \eqref{equation-1-FLRW} and \eqref{equation-2-FLRW} can be interpreted as the $00$- and $ii$-components of (tensorial) Einstein field equation, respectively.
\section{Exact de Sitter solution} \label{sec3}
Given the field equations of the SGZ gravity model derived above, we now would like to look for their exact de Sitter solution. With the suggestion from the previous investigations on fourth-order gravities \cite{Do:2023yvg,Pham:2024fub,preprint,Do:2020vdc,Toporensky:2006kc,Barrow:2005qv,Barrow:2006xb,Barrow:2009gx}, we would like to take an ansatz for the scale factor such as
\begin{equation} \label{ansatz}
\alpha(t) =\zeta t,
\end{equation}
where $\zeta$ is an undetermined constant, while $t$ is nothing but the cosmic time. It is apparent that this ansatz is chosen since it exactly corresponds  to the de Sitter form of scale factor, i.e.,
\begin{equation}
a(t) =e^{\alpha(t)} = e^{\zeta t}.
\end{equation}
Of course, one can choose another ansatz such as $\alpha(t)=\zeta t+ \zeta_0$ with $\zeta_0$ being another undetermined constant. In this case, the scale factor will take the following form,
\begin{equation}
a(t) =e^{\alpha(t)} = e^{\zeta_0} e^{\zeta t}.
\end{equation}
However, the factor $e^{\zeta_0}$ can be omitted by taking a rescaling such as $(x,y,z) \to e^{-\zeta_0} (x,y,z)$. This means that $\zeta_0$ can be ignored in the chosen ansatz for simplicity. Remarkably, the ansatz \eqref{ansatz} will significantly reduce the complicated field equations to very simple ones by eliminating higher-than-two-derivative terms.

As a result, Eqs. \eqref{equation-1-FLRW} and \eqref{equation-2-FLRW} all lead to the same algebraic  equation of $\zeta$,
\begin{equation} \label{zeta-eta-4}
12 \bar\gamma  \zeta ^6-1 =0,
\end{equation}
whose solution is nothing but the de Sitter one with 
\begin{equation} \label{sol-of-zeta-FLRW}
a(t) = e^{\zeta t}, \quad \zeta = \left(\frac{1}{12\bar\gamma}\right)^{\frac{1}{6}}.
\end{equation}

Interestingly, this solution is identical to one found in the EGZ gravity \cite{CamposDelgado:2024jst,preprint}, meaning that  the Starobinsky $R^2$ term does not contribute to the value of the found de Sitter solution.  It is noted that the pure Starobinsky gravity model does not admit an exact de Sitter solution. Instead, it admits a quasi-de Sitter one as an attractor \cite{Starobinsky:1980te}.

The parameter $\bar\gamma$ must be positive definite in order to ensure the existence of real $\zeta$. The positivity of $\bar\gamma$ implies that $\gamma$ must be positive definite, too, consistent with the previous studies on the de Sitter solution in the EGZ gravity \cite{CamposDelgado:2024jst,preprint}. Furthermore, if the obtained de Sitter solution represents an inflationary phase of early universe with  $\zeta \gg 1$,  then $\bar\gamma$ must be much smaller than one. On the other hand, if the obtained de Sitter solution represents an accelerated expansion of late-time universe with $0<\zeta \leq 1 $, then it will require that $\bar\gamma \geq 1/12$. 
\section{Dynamical system} \label{sec4}
In this section, the stability of the obtained de Sitter solution will be investigated by using the dynamical system method \cite{Do:2023yvg,Pham:2024fub,preprint,Do:2020vdc,Toporensky:2006kc,Barrow:2005qv,Barrow:2006xb,Barrow:2009gx}.  To do this, the field equations must be transformed into the corresponding dynamical system. Therefore, we first introduce suitable dimensionless dynamical variables with the hint from previous studies in fourth-order gravities \cite{Do:2023yvg,Pham:2024fub,preprint,Do:2020vdc,Barrow:2005qv,Barrow:2006xb,Barrow:2009gx}
 \begin{align}
& B=\frac{1}{\dot\alpha^2},\\
&Q=\frac{\ddot\alpha}{\dot\alpha^2},\\
& Q_2 =\frac{\alpha^{(3)}}{\dot\alpha^3}.
 \end{align}
 As a result, the corresponding dynamical system of the SGZ gravity is given by 
 \begin{align}
  \label{Dyn-1}
 B' &= -2QB,\\
  \label{Dyn-2}
 Q' &=Q_2 -2Q^2,\\
 \label{Dyn-3}
 Q_2 '&= \frac{\alpha^{(4)}}{\dot\alpha^4} -3Q Q_2,
 \end{align}
 with the help of a result coming from Eq. \eqref{equation-2-FLRW},
  \begin{align} \label{Dyn-4-1}
\frac{\alpha^{(4)}}{\dot\alpha^4} = &~ \frac{-1}{12\hat\gamma+ 4\bar\gamma \left(3Q^2 +12 Q +11  \right) } \left\{ 2B^3 Q+3B^3 +6\hat\gamma \left(12 B^2 Q_2 +9B^2 Q^3 +18 B^2 Q \right) \right. \nonumber\\
&\left. + \bar\gamma \left[24 \left( Q+ 2 \right)  Q_2^2 + \left(264 Q^2 +640 Q +264  \right) Q_2  + 47 Q^4 +636  Q^3 \right. \right. \nonumber\\
&\left. \left. +1206  Q^2 + 300 Q -36  \right] \right\}.
\end{align}
Here, it appears that $B' =dB/d\tau$ and so on, with $\tau$ is nothing but a dynamical time variable defined as $\tau =\int \dot\alpha dt$. In addition,  the Friedmann equation \eqref{equation-1-FLRW} can be rewritten in terms of the dynamical variables as follows
 \begin{align} \label{Dyn-5}
 & B^3 +6 \hat\gamma\left(2B^2 Q_2 -B^2 Q^2 +6B^2 Q \right) \nonumber\\
&+  \bar\gamma \left[ 4\left( 3Q^2 +12 Q +11  \right) Q_2 - 3 Q^4 + 28 Q^3 +138 Q^2 + 132 Q -12  \right] =0.
 \end{align}
 This equation plays as a constraint equation, which all found fixed points must obey.
 \subsection{Fixed points}
 Given the above dynamical system, we are going to figure out its fixed point(s), which should be equivalent to the obtained de Sitter solution with $B\neq 0$. Mathematically, fixed points satisfy the following set of equations, \cite{Do:2023yvg,Pham:2024fub,preprint,Do:2020vdc,Toporensky:2006kc,Barrow:2005qv,Barrow:2006xb,Barrow:2009gx}, 
 \begin{equation}
B'=Q'=Q_2' =0.
\end{equation}
Consequently, we have, according to Eqs. \eqref{Dyn-1}, \eqref{Dyn-2}, and \eqref{Dyn-3}, that
 \begin{equation}
 Q=Q_2=\frac{\alpha^{(4)}}{\dot\alpha^4}=0,
 \end{equation}
 for $B \neq 0$. 
 As a result, both equations, ${\alpha^{(4)}}/{\dot\alpha^4}=0$ and  Eq. \eqref{Dyn-5}, lead to the same equation of $B$,
 \begin{equation}
 B^3 -12 \bar\gamma  =0.
 \end{equation}
 As a result, an exact solution of this equation can be integrated out, up to an integration constant, to be
 \begin{equation}
 \alpha(t)= \zeta t,
 \end{equation}
with $\zeta$ has been defined in Eq. \eqref{sol-of-zeta-FLRW}. This result  confirms our expectation that this fixed point is indeed equivalent to the de Sitter solution found in the previous section. 

It is noted that the obtained fixed point is not the only solution to the set of equations, $B'=Q'=Q_2' =0$. Indeed, one can easily identify another fixed point having $B=0$, according to the equation \eqref{Dyn-1}.  Hence, the equation $Q_2'=0$ implies the corresponding equation of $Q$, 
\begin{equation} \label{equation-of-Q}
21Q^4 +124Q^3 +226Q^2 +132Q-12=0,
\end{equation}
with the help of the associated relation, $Q_2 = 2Q^2$, derived from the remaining equation $Q'=0$ along with a requirement, $Q \neq 0$. As a result, two real solutions, $Q \simeq -3.1199$ and $Q\simeq 0.0796$, can be found from Eq. \eqref{equation-of-Q}.  Unfortunately, this fixed point is not our desired solution since it is not equivalent to the de Sitter solution found in the previous section. For convenience, we will call it the non-de Sitter fixed point, similar to our previous paper \cite{preprint}. Remarkably, this non-de Sitter fixed point will play a complementary role in describing the full behavior of the dynamical system, as illustrated in Fig. \ref{fig1}.
 \subsection{Stability of the de Sitter fixed point}
So far, the fixed point, which is absolutely equivalent to the de Sitter solution of the SGZ gravity, has been identified explicitly from the dynamical system. Now, we would like to examine its stability and attractor properties.  Similar to Ref. \cite{preprint}, we first perturb the autonomous equations \eqref{Dyn-1}, \eqref{Dyn-2}, and \eqref{Dyn-3} around the obtained fixed point,
\begin{align}
\label{pert-1}
\delta B' &=-2B\delta Q,\\
\label{pert-2}
\delta Q' &= \delta Q_2,\\
\label{pert-3}
\delta Q_2' &= \frac{-1}{12\hat\gamma+ 44 \bar\gamma} \left( 2B^3 \delta Q + 9B^2 \delta B +72\hat\gamma B^2 \delta Q_2 +48\hat\gamma B^2 \delta Q+ 264\bar\gamma \delta Q_2 +300\bar\gamma \delta Q \right).
\end{align}
 Furthermore, Eq. \eqref{pert-3} can be simplified as follows
\begin{align}\label{pert-5}
\delta Q_2' = \frac{-1}{12\hat\gamma+ 44\bar\gamma}  \left( 2B^3 \delta Q -60\hat\gamma B^2 \delta Q -96\bar\gamma \delta Q +36\hat\gamma B^2 \delta Q_2   + 132 \bar\gamma \delta Q_2  \right),
\end{align}
with the help of perturbations derived from Eq. \eqref{Dyn-5}.
By taking exponential perturbations \cite{preprint}, 
\begin{align}
\delta B &= C_B \exp[\mu\tau], \\
\delta Q &= C_Q \exp[\mu\tau],\\
\delta Q_2 &= C_{Q_2}  \exp[\mu\tau],
\end{align} the perturbed equations \eqref{pert-1}, \eqref{pert-2}, and \eqref{pert-5}, will reduce to the following homogeneous system of linear equations, which can be written in a matrix form as
  \begin{align} \label{stability-equation}
& {\cal M}\left( {\begin{array}{*{20}c}
   C_B  \\
   C_Q  \\
   C_{Q_2} \\
 \end{array} } \right) \nonumber\\
 &\equiv \left[ {\begin{array}{*{20}c}
   {\mu} & {2B} & {0 }   \\
   { 0} & {\mu} & {-1 }  \\
     {0  } & {\frac{1}{12\hat\gamma+ 44\bar\gamma } \left( 2B^3 -60\hat\gamma B^2  -96\bar\gamma \right) } & {  \mu+ \frac{1}{12\hat\gamma+ 44\bar\gamma } \left(36\hat\gamma B^2 +132\bar\gamma \right) }  \\
 \end{array} } \right]  \left( {\begin{array}{*{20}c}
   C_B  \\
   C_Q  \\
   C_{Q_2} \\
 \end{array} } \right) = 0.
\end{align}
Mathematically, this homogeneous system admits non-trivial solutions, in which at least one of three $C_i$ ($i=B,~Q,$ and $Q_2$) is not equal to zero, if and only if
\begin{equation}
\det {\cal M} =0.
\end{equation}
As a result, this equation can be expanded, thanks to the value of the fixed point, $B =\left(12 \bar\gamma \right)^{\frac{1}{3}}$, to be a cubic equation of $\mu$ given by 
\begin{equation} \label{quadratic-equation}
\mu  \left(a_2 \mu^2+a_1 \mu +a_0  \right) =0,
\end{equation}
with
\begin{align}
a_2 &= 3\hat\gamma +11\bar\gamma,\\
a_1&= 18 \hat\gamma \left(18  \bar\gamma^2 \right)^{\frac{1}{3}} +33\bar\gamma,\\
a_0&= -30\hat\gamma \left(18  \bar\gamma^2 \right)^{\frac{1}{3}} -18\bar\gamma.
\end{align}
If $\hat\gamma=0$, or equivalently the Starobinsky $R^2$ term is turned off, then Eq. \eqref{quadratic-equation} will reduce to a simple one,
\begin{equation}
\bar\gamma \mu  \left(11 \mu^2+33 \mu -18  \right) =0,
\end{equation}
which is nothing but that derived in our previous paper on the EGZ gravity \cite{preprint}. More interestingly, this equation always admits one positive root $\mu = \mu_3 \equiv -3 \left(11-\sqrt{209} \right)/22 >0$, which is clearly independent of $\bar\gamma$. This clearly implies that the de Sitter solution of the EGZ gravity is always unstable, regardless of the value of $\bar\gamma$ \cite{preprint}. 

\begin{figure}\centering
	  \includegraphics[scale=0.7]{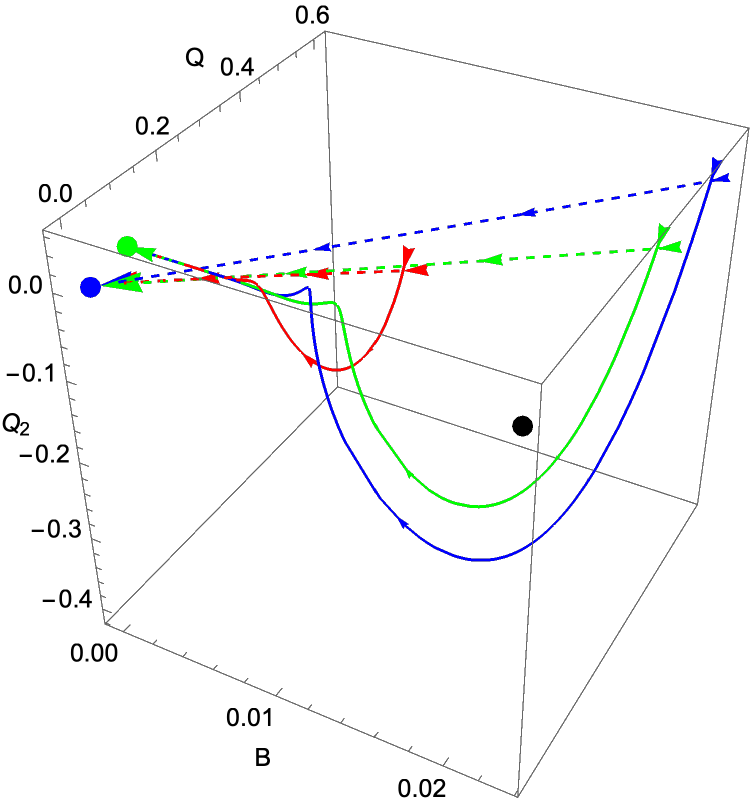}
	
\caption{\it The de Sitter fixed point displayed as a black point with $(B,Q,Q_2) \simeq (0.023,0,0)$ acts as a repeller since all trajectories tend to repel it. In particular, they tend to converge to either the non-de Sitter fixed point with $(B,Q,Q_2) \simeq (0,0.0796,0.0127)$  (solid curves and a green point) or a trivial (unphysical) fixed point with $(B,Q,Q_2) = (0,0,0)$ (dashed curves and a blue point), depending on the value of $\hat\gamma$. The parameters have been chosen as  $\hat\gamma =10^{-5}$, $\bar\gamma=10^{-6}$, and $M_p=1$ for the solid curves and green point; while $\hat\gamma =10^{4}$, $\bar\gamma=10^{-6}$, and $M_p=1$ for the dashed curves and blue point. Different colors of trajectories correspond to different initial conditions. }
\label{fig1}
\end{figure}

Now, thanks to the presence of the Starobinsky $R^2$ term with the non-vanishing $\hat\gamma$ the GZ term can play its role, in terms of $\bar\gamma$, in the stability of the de Sitter solution of the SGZ gravity. As shown above, $\hat\gamma$ is always positive by its definition shown in Eq. \eqref{def-gamma}, while $\bar\gamma$ must be positive definite due to the existence of de Sitter solution as pointed out in Eq.  \eqref{sol-of-zeta-FLRW}. Consequently, $a_2$ turns out to be positive definite, while $a_0$ is apparently negative definite. Hence, it can be easily concluded, without explicit solutions, that Eq. \eqref{quadratic-equation} always admits at least one positive root $\mu >0$. In other words, the de Sitter fixed point of the SGZ gravity is always unstable. Furthermore, it is confirmed through numerical calculations that this de Sitter fixed point is indeed a repeller of the dynamical system (see Fig. \ref{fig1} for details). All these results imply, according to the discussions in Refs.  \cite{Elizalde:2014xva,Pozdeeva:2019agu,Vernov:2021hxo}, that the SGZ gravity would be more compatible with the inflationary phase of early universe rather than the accelerated expansion of late-time universe. In particular, the instability of de Sitter solution would make the SGZ gravity free of the so-called eternal inflation and therefore a multiverse scenario \cite{Guth:2007ng}, which seems to be resolved only via a graceful exit \cite{Brustein:1994kw}.

\section{Conclusions}\label{final}
We have investigated whether the SGZ gravity proposed recently in Ref. \cite{Toyama:2024ugg} admits a stable de Sitter solution. Our investigation is based on two effective calculation methods: the first one is the EL equations for deriving the field equations and the other is the dynamical system for examining the stability of the de Sitter solution. Remarkably, these two methods have been used extensively in our previous papers \cite{Do:2023yvg,Pham:2024fub,preprint,Do:2020vdc}. It is apparent that our results turn out to be consistent with Ref. \cite{Toyama:2024ugg}. As a result, the SGZ gravity has been shown to admit an unstable de Sitter solution. Furthermore, it has been shown that although the Starobinsky $R^2$ term does not contribute to the value of the obtained de Sitter solution but it does affect on the stability and attractive properties of the de Sitter fixed point.  
This important result suggests that the SGZ gravity would only be relevant to the inflationary phase of early universe. Our present paper together with the previous on the EGZ gravity \cite{preprint} point out that the GZ term could play non-trivial roles in understanding the inflationary phase of early universe. Other cosmological implications of this term should be considered extensively. Hence, related investigations of the SGZ gravity during the inflationary phase will be our future studies. For now, we hope that the present work could be relevant to the stability analysis of other fourth-order gravities, e.g., the Einsteinian cubic gravity \cite{Bueno:2016xff,Arciniega:2018fxj}. To end this section, we would like to emphasize that our stability analysis provide a strong evidence that the SGZ gravity can be safely viewed as a promising modification of the Starobinsky model, which might help us to resolve its tension with the latest ACT data \cite{AtacamaCosmologyTelescope:2025nti,Addazi:2025qra,Bianchi:2025tyl,Ketov:2025cqg}. This claim is supported by the fact that the SGZ gravity does admit the unstable de Sitter inflationary solution, and therefore avoids the eternal inflation problem, similar to the pure Starobinsky model.  Without a doubt, this is the most important result of our present paper.
\begin{acknowledgments}
  The author would like to thank an anonymous referee very much for his/her very useful comments and suggestions.  This study is funded by the Vietnam National Foundation for Science and Technology Development (NAFOSTED) under grant number 103.01-2023.50. The author would like to thank Prof. Phung V. Dong very much for his support.
\end{acknowledgments}


\begin{thebibliography}{99} 
\bibitem{Starobinsky:1980te}
A.~A.~Starobinsky,
A new type of isotropic cosmological models without singularity,
Phys. Lett. B {\bf 91}, 99 (1980).
  

 \bibitem{Guth:1980zm} 
  A.~H.~Guth,
  The inflationary universe: A possible solution to the horizon and flatness problems,
  Phys.\ Rev.\ D {\bf 23}, 347 (1981).
  
\bibitem{Linde:1981mu} 
  A.~D.~Linde,
  A new inflationary universe scenario: A possible solution of the horizon, flatness, homogeneity, isotropy and primordial monopole problems,
  Phys.\ Lett.\  {\bf 108B}, 389 (1982).
  
\bibitem{Linde:1983gd} 
  A.~D.~Linde,
  Chaotic inflation,
  Phys.\ Lett.\  {\bf 129B}, 177 (1983). 
  
\bibitem{Akrami:2018odb}
Y.~Akrami \textit{et al.} [Planck],
Planck 2018 results. X. Constraints on inflation,
Astron. Astrophys. \textbf{641}, A10 (2020)
[arXiv:1807.06211].


     
\bibitem{Whitt:1984pd}
B.~Whitt, Fourth order gravity as general relativity plus matter,
Phys. Lett. B \textbf{145}, 176 (1984).

\bibitem{Maeda:1987xf}
K.~i.~Maeda, Inflation as a transient attractor in $R^2$ cosmology,
Phys. Rev. D \textbf{37}, 858 (1988).

\bibitem{Barrow:1988xh}
J.~D.~Barrow and S.~Cotsakis, Inflation and the conformal structure of higher order gravity theories, 
Phys. Lett. B \textbf{214}, 515 (1988).

 
\bibitem{Carroll:2004de}
S.~M.~Carroll, A.~De Felice, V.~Duvvuri, D.~A.~Easson, M.~Trodden, and M.~S.~Turner,
The cosmology of generalized modified gravity models,
Phys. Rev. D \textbf{71}, 063513 (2005)
[astro-ph/0410031].

\bibitem{Nojiri:2010wj}
S.~Nojiri and S.~D.~Odintsov,
Unified cosmic history in modified gravity: from F(R) theory to Lorentz non-invariant models,
Phys. Rept. \textbf{505}, 59 (2011)
[arXiv:1011.0544].

\bibitem{Nojiri:2017ncd}
S.~Nojiri, S.~D.~Odintsov, and V.~K.~Oikonomou,
Modified gravity theories on a nutshell: Inflation, bounce and late-time evolution,
Phys. Rept. \textbf{692}, 1 (2017)
[arXiv:1705.11098].

\bibitem{Stelle:1976gc}
K.~S.~Stelle,
Renormalization of higher derivative quantum gravity,
Phys. Rev. D \textbf{16}, 953 (1977).

  


\bibitem{Woodard:2015zca}
R.~P.~Woodard,
Ostrogradsky's theorem on Hamiltonian instability,
Scholarpedia \textbf{10},  32243 (2015)
[arXiv:1506.02210].
  
  
\bibitem{Schmidt:2006jt}
H.~J.~Schmidt,
Fourth order gravity: Equations, history, and applications to cosmology,
eConf \textbf{C0602061}, 12 (2006)
[gr-qc/0602017].

\bibitem{Alvarez-Gaume:2015rwa}
L.~Alvarez-Gaume, A.~Kehagias, C.~Kounnas, D.~L{\"u}st, and A.~Riotto,
Aspects of quadratic gravity,
Fortsch. Phys. \textbf{64},  176 (2016)
[arXiv:1505.07657].


\bibitem{Salvio:2018crh}
A.~Salvio,
Quadratic gravity,
Front. in Phys. \textbf{6}, 77 (2018)
[arXiv:1804.09944].

\bibitem{Ketov:2025nkr}
S.~V.~Ketov,
On Legacy of Starobinsky Inflation,
arXiv:2501.06451.


\bibitem{AtacamaCosmologyTelescope:2025nti}
E.~Calabrese \textit{et al.} [Atacama Cosmology Telescope],
The Atacama Cosmology Telescope: DR6 constraints on extended cosmological models, J. Cosmol. Astropart. Phys. \textbf{11}, 063 (2025)
[arXiv:2503.14454].



\bibitem{Myrzakulov:2014hca}
R.~Myrzakulov, S.~Odintsov, and L.~Sebastiani,
Inflationary universe from higher-derivative quantum gravity,
Phys. Rev. D \textbf{91}, 083529 (2015)
[arXiv:1412.1073].

\bibitem{Cano:2020oaa}
P.~A.~Cano, K.~Fransen, and T.~Hertog,
Novel higher-curvature variations of $R^2$ inflation,
Phys. Rev. D \textbf{103},  103531 (2021)
[arXiv:2011.13933].

\bibitem{Rodrigues-da-Silva:2021jab}
G.~Rodrigues-da-Silva, J.~Bezerra-Sobrinho, and L.~G.~Medeiros,
Higher-order extension of Starobinsky inflation: Initial conditions, slow-roll regime, and reheating phase,
Phys. Rev. D \textbf{105},  063504 (2022)
[arXiv:2110.15502].

\bibitem{Ivanov:2021chn}
V.~R.~Ivanov, S.~V.~Ketov, E.~O.~Pozdeeva, and S.~Y.~Vernov,
Analytic extensions of Starobinsky model of inflation,
J. Cosmol. Astropart. Phys. \textbf{03},  058 (2022)
[arXiv:2111.09058].

\bibitem{Gialamas:2023lxj}
I.~D.~Gialamas and K.~Tamvakis,
Bimetric Starobinsky model,
Phys. Rev. D \textbf{108},  104023 (2023)
[arXiv:2307.05673].

\bibitem{Modak:2022gol}
T.~Modak, L.~R\"over, B.~M.~Sch\"afer, B.~Schosser, and T.~Plehn,
Cornering extended Starobinsky inflation with CMB and SKA, SciPost Phys. {\bf 15}, 047 (2023)
[arXiv:2210.05698].


\bibitem{Ketov:2022lhx}
S.~V.~Ketov,
Starobinsky\textendash{}Bel\textendash{}Robinson Gravity,
Universe \textbf{8},  351 (2022)
[arXiv:2205.13172].

\bibitem{Ketov:2022zhp}
S.~V.~Ketov, E.~O.~Pozdeeva, and S.~Y.~Vernov,
On the superstring-inspired quantum correction to the Starobinsky model of inflation,
 J. Cosmol. Astropart. Phys.  \textbf{12}, 032 (2022)
[arXiv:2211.01546].


\bibitem{Do:2023yvg}
T.~Q.~Do, D.~H.~Nguyen, and T.~M.~Pham,
Stability investigations of isotropic and anisotropic exponential inflation in the Starobinsky{\textendash}Bel{\textendash}Robinson gravity,
Int. J. Mod. Phys. D \textbf{32},  2350087 (2023)
[arXiv:2303.17283].

\bibitem{Pham:2024fub}
T.~M.~Pham, D.~H.~Nguyen, T.~Q.~Do and W.~F.~Kao,
Stability investigations of de Sitter inflationary solutions in power-law extensions of the Starobinsky model,
Eur. Phys. J. C \textbf{84}, 729 (2024)
[arXiv:2403.02623].

\bibitem{Toyama:2024ugg}
S.~Toyama and S.~V.~Ketov,
Starobinsky inflation beyond the leading order,
Phys. Rev. D \textbf{110},  6 (2024)
[arXiv:2407.21349].

\bibitem{Asorey:2024oxw}
M.~Asorey, F.~Ezquerro, and M.~Pardina,
Stability of cosmological singularity-free solutions in quadratic gravity,
Phys. Rev. D \textbf{111},  064020 (2025)
[arXiv:2412.10111].


\bibitem{Addazi:2025qra}
A.~Addazi, Y.~Aldabergenov, and S.~V.~Ketov,
Curvature corrections to Starobinsky inflation can explain the ACT results,
Phys. Lett. B \textbf{869}, 139883 (2025)
[arXiv:2505.10305].

\bibitem{Bianchi:2025tyl}
E.~Bianchi and M.~Gamonal,
Precision predictions of Starobinsky inflation with self-consistent Weyl-squared corrections,
Phys. Rev. D \textbf{112}, 124006 (2025)
[arXiv:2506.10081].

\bibitem{Ketov:2025cqg}
S.~V.~Ketov, E.~O.~Pozdeeva, and S.~Y.~Vernov,
Inflation in F(R) gravity models revisited after ACT,
 J. Cosmol. Astropart. Phys. \textbf{12}, 040 (2025)
[arXiv:2508.08927].



\bibitem{CamposDelgado:2024jst}
R.~Campos Delgado and S.~V.~Ketov,
Einstein-Grisaru-Zanon gravity,
Phys. Lett. B \textbf{855}, 138811 (2024)
[arXiv:2405.03925].



\bibitem{Grisaru:1986vi}
M.~T.~Grisaru and D.~Zanon,
$\sigma$ model superstring corrections to the Einstein-Hilbert action,
Phys. Lett. B \textbf{177}, 347 (1986).


 \bibitem{preprint}
 T. Q. Do, Stability analysis of de Sitter solution in the Einstein-Grisaru-Zanon gravity using the dynamical system method, Int. J. Geom. Meth. Mod. Phys. (2026) 2650168. [DOI: 10.1142/S0219887826501689]
 


\bibitem{Bahamonde:2017ize}
S.~Bahamonde, C.~G.~B\"ohmer, S.~Carloni, E.~J.~Copeland, W.~Fang, and N.~Tamanini,
Dynamical systems applied to cosmology: dark energy and modified gravity,
Phys. Rept. \textbf{775-777}, 1 (2018)
[arXiv:1712.03107].

   
\bibitem{Do:2020vdc}
T.~Q.~Do,
No-go theorem for inflation in Ricci-inverse gravity,
Eur. Phys. J. C \textbf{81}, 431 (2021)
[arXiv:2009.06306].


\bibitem{Elizalde:2014xva}
E.~Elizalde, S.~D.~Odintsov, E.~O.~Pozdeeva, and S.~Y.~Vernov,
Renormalization-group improved inflationary scalar electrodynamics and SU(5) scenarios confronted with Planck 2013 and BICEP2 results,
Phys. Rev. D \textbf{90}, 084001 (2014)
[arXiv:1408.1285].

\bibitem{Pozdeeva:2019agu}
E.~O.~Pozdeeva, M.~Sami, A.~V.~Toporensky, and S.~Y.~Vernov,
Stability analysis of de Sitter solutions in models with the Gauss-Bonnet term,
Phys. Rev. D \textbf{100}, 083527 (2019)
[arXiv:1905.05085].

\bibitem{Vernov:2021hxo}
S.~Vernov and E.~Pozdeeva,
De Sitter Solutions in Einstein\textendash{}Gauss\textendash{}Bonnet Gravity,
Universe \textbf{7}, 149 (2021)
[arXiv:2104.11111].

\bibitem{Toporensky:2006kc}
A.~V.~Toporensky and P.~V.~Tretyakov,
De Sitter stability in quadratic gravity,
Int. J. Mod. Phys. D \textbf{16}, 1075 (2007)
[gr-qc/0611068].

\bibitem{Kamenshchik:2024kay}
A.~Y.~Kamenshchik, E.~O.~Pozdeeva, A.~Tribolet, A.~Tronconi, G.~Venturi, and S.~Y.~Vernov,
Superpotential method and the amplification of inflationary perturbations,
Phys. Rev. D \textbf{110},  104011 (2024)
[arXiv:2406.19762].



\bibitem{Guth:2007ng}
A.~H.~Guth,
Eternal inflation and its implications,
J. Phys. A \textbf{40}, 6811 (2007)
[hep-th/0702178].

\bibitem{Brustein:1994kw}
R.~Brustein and G.~Veneziano,
The Graceful exit problem in string cosmology,
Phys. Lett. B \textbf{329}, 429 (1994)
[hep-th/9403060].



\bibitem{Kao:1991zz}
W.~F.~Kao and U.~L.~Pen,
Generalized Friedmann-Robertson-Walker metric and redundancy in the generalized Einstein equations, Phys. Rev. D \textbf{44}, 3974 (1991).


\bibitem{Barrow:2005qv} 
  J.~D.~Barrow and S.~Hervik,
  Anisotropically inflating universes,
  Phys.\ Rev.\ D {\bf 73}, 023007 (2006)
  [gr-qc/0511127].


\bibitem{Barrow:2006xb} 
  J.~D.~Barrow and S.~Hervik,
  On the evolution of universes in quadratic theories of gravity,
  Phys.\ Rev.\ D {\bf 74}, 124017 (2006)
  [gr-qc/0610013].
  
\bibitem{Barrow:2009gx} 
  J.~D.~Barrow and S.~Hervik,
  Simple types of anisotropic inflation,
  Phys.\ Rev.\ D {\bf 81}, 023513 (2010)
  [arXiv:0911.3805].
   

\bibitem{Bueno:2016xff}
P.~Bueno and P.~A.~Cano,
Einsteinian cubic gravity,
Phys. Rev. D \textbf{94}, 104005 (2016)
[arXiv:1607.06463].

\bibitem{Arciniega:2018fxj}
G.~Arciniega, J.~D.~Edelstein, and L.~G.~Jaime,
Towards geometric inflation: the cubic case,
Phys. Lett. B \textbf{802}, 135272 (2020)
[arXiv:1810.08166].

\end{thebibliography}
\end{document}